\documentclass[a4paper]{jpconf}
\usepackage{graphicx}
\usepackage{iopams}
\usepackage{subfig}
\usepackage{sidecap}
\usepackage{ dsfont }
\usepackage{amsmath}
\usepackage{placeins}
\usepackage{xcolor}

\usepackage{hyperref}
\hypersetup{
colorlinks=true,
citecolor=blue,
citebordercolor=red,
linktoc=all,
linkcolor=blue}

\newcommand{\be}{\begin{equation}}
\newcommand{\ee}{\end{equation}}
\newcommand{\bea}{\begin{eqnarray}}
\newcommand{\eea}{\end{eqnarray}}

\graphicspath{{./fig/}}
\begin{document}

\title{Exploration of the phase diagram and the thermodynamic properties of QCD at finite temperature and chemical potential with the PNJL effective model}
\author{M.Motta$^{1,2}$, W. M. Alberico$^{1,2}$, A. Beraudo$^1$, P. Costa$^3$ and R. Stiele$^{1,4}$}
\address{$^1$ INFN -- Sezione di Torino, Via Pietro Giuria 1, I-10125 Torino, Italy}
\address{$^2$ Dipartimento di Fisica, Universit\`a degli Studi di Torino, Via Pietro Giuria 1, I-10125 Torino, Italy}
\address{$^3$ CFisUC, Department of Physics, University of Coimbra, P-3004 - 516  Coimbra, Portugal}
\address{$^4$ Univ.~Lyon, ENS de Lyon, Univ Claude Bernard Lyon 1, CNRS, F-69342 Lyon, France}

\begin{abstract}
The QCD transition from a hadronic to a quark-gluon plasma phase is a cross-over at vanishing/small baryo-chemical potential, while at higher chemical potentials it is argued that it becomes of first order, ending with a critical end-point. The present goal is the determination of the critical line and, possibly, the recognition of the critical endpoint.  For this purpose, the effective model (Nambu-Jona-Lasinio model with Polyakov Loop) with 2+1 flavours is used. In Mean Field Approximation it is possible to obtain all thermodynamic quantities (pressure, entropy, energy and quark density) and their fluctuations, in particular, the generalized quark-number susceptibilities. We use the net baryon-number fluctuations and net-strangeness fluctuations to identify the critical line and the order of the phase transition in the chemical potential - temperature plane.
\end{abstract}

\section{Introduction}
The study of the QCD phase diagram and thermodynamics has been a subject of intense investigation in recent years. The lattice-QCD simulations provide very good results at zero or small chemical potential, but for phenomenological applications it is important to explore the region at high chemical potential \cite{Lattice}. For this purpose a different approach is needed, e.g. an effective field theory with two main characteristics: chiral-symmetry breaking/restoration and a modeling of quark confinement. The NJL model (see \cite{Kl}) with Polyakov loop corrections in 2+1 flavors (PNJL) satisfies this feature. \\
In this framework there are 3 different chemical potentials associated to conserved charges: baryon-number($\mu_B$), electric charge($\mu_Q$) and strangeness ($\mu_S$). To perform the calculations the values of two of these chemical potentials are fixed by experimental constraints and the third is kept free. In this work we explore the scenarios with $\mu_Q=\mu_S=0$ and  $\mu_Q=0$ and $\mu_S=1/3\mu_B$ (for more details see \cite{nuovo_1},\cite{nuovo_2}).  
 We show how the PNJL effective model with fluctuations of net-baryon-number can be used to locate the critical-end-point.

\section{PNJL 2+1 flavour Lagrangian}
The Lagrangian of PNJL effective field model reads: 
\begin{equation}
\label{eq:lagrangian}
\begin{split}
\mathcal{L}&_{PNJL} =\bar{q}(i \gamma^\mu D_\mu -\hat{m})q + G[(\bar{q}\vec{\tau}q)^2 + (\bar{q}i\gamma^5\vec{\tau}q)^2]+\\&+K\{\text{det}[\bar{q}(1+\gamma^5)q]+\text{det}[\bar{q}(1-\gamma^5)q]\}-\mathcal{U}(\Phi[A],\bar{\Phi}[A],T)-i\hat\mu\bar{q}\gamma^0q
\end{split}
\end{equation}
Here $D^\mu=\partial^\mu-iA^\mu$, $A^\mu=\delta^\mu_0A^0$, the fields $\Phi$ and $\bar{\Phi}$ are Polyakov fields defined as:
\begin{equation}
\Phi\equiv\frac{1}{N_c}\text{Tr}\langle\langle L \rangle\rangle \qquad \bar\Phi\equiv\frac{1}{N_c}\text{Tr}\langle\langle L^\dag \rangle\rangle,
\end{equation}
where $L$ is the Polyakov loop derived from the gauge field $A_4$, after Wick rotation:
\begin{equation}
L(\vec x)\equiv\mathcal{P}\Big\{i\int_0^\beta d\tau A_4(\tau,\vec x)\Big\}.
\end{equation}

\section{Thermodynamics in Mean Field Approximation}
From Eq. (\ref{eq:lagrangian}) one obtains the Thermodynamic Potential per unit volume($\omega=\Omega/V$) in Mean Field Approximation (MFA):
\begin{equation}
\begin{split}
&\omega(\Phi,\bar\Phi,T,M_i,\mu_i)=\mathcal{U}(\Phi,\bar\Phi,T)+G\sum_{i=u,d,s}\varphi_i^2+K\varphi_u\varphi_d\varphi_s+\\-&2N_c\sum_{i=u,d,s}\int^\Lambda\frac{d^3p}{(2\pi)^3}E_i
-2T\int^\Lambda\frac{d^3p}{(2\pi)^3}\big\{z^{i+}_\Phi(E_i,\mu)+z^{i-}_\Phi(E_i,\mu)\}
\end{split}
\end{equation}
where the functions $z^{i\pm}$ are 
\begin{equation}
z^{i+}_\Phi(E_i,\mu_i)\equiv\ln[1+N_c(\Phi+\bar\Phi\e^{-\beta(E_i-\mu_i)})\e^{-\beta(E_i-\mu_i)}+\e^{-3\beta(E_i-\mu_i)}]
\end{equation}
\begin{equation}
z^{i-}_\Phi(E_i,\mu_i)\equiv\ln[1+N_c(\bar\Phi+\Phi\e^{-\beta(E_i+\mu_i)})\e^{-\beta(E_i+\mu_i)}+\e^{-3\beta(E_i+\mu_i)}]
\end{equation}
and $\varphi\equiv\langle \bar q_i q_i\rangle \ (i=u,d,s)$ are the quark chiral condensates. The quark masses are related to the chiral condensates by the Gap Equation Eq.\ref{eq:MGE} discussed in the following.

The effective potential $\mathcal{U}(\Phi,\bar{\Phi};T)$ is real (it appears in the Grand Canonical Potential), it must respect $\mathds{Z}_3$ symmetry and depends explicitly on temperature, allowing one to describe the transition from confinement ($\Phi\rightarrow 0$) to deconfinement ($\Phi\rightarrow 1$). The potential proposed in \cite{Ratti} satisfies these requests.
\begin{equation}
\frac{\mathcal{U}(\Phi,\bar\Phi;T)}{T^4}=-\frac{a(T)}{2}\bar\Phi\Phi+b(T)\ln[1-6\bar\Phi\Phi+4(\Phi^3+\bar\Phi^3)-3(\bar\Phi\Phi)^2]
\end{equation}
where
\begin{equation}
a(T)=a_0+a_1\bigg(\frac{T_0}{T}\bigg)+a_2\bigg(\frac{T_0}{T}\bigg)^2 \qquad  \qquad b(T)=b_3\bigg(\frac{T_0}{T}\bigg)^3
\end{equation}
By minimizing the thermodynamic potential one gets the mean field equations (MFE):
\begin{equation}
\frac{\partial \omega}{\partial \varphi_i}=0 \qquad \frac{\partial \omega}{\partial \Phi}=0 \qquad \frac{\partial \omega}{\partial \bar\Phi}=0
\end{equation}
from which one derives all the thermodynamic quantities of interest, e.g: the pressure and the density of particle of species $i$ 
\begin{equation}
	P(T,\mu_i)=-\omega(T,\mu_i), \qquad n_i(T,\mu_i)\equiv-\frac{\partial \omega(T,\mu_i)}{\partial \mu_i}\Big |_T
\end{equation}

The PNJL Lagrangian is chiral symmetric if $m_i=0$ . For non-vanishing current mass chiral symmetry is explicitly broken, although this breaking is small. Moreover, at low temperature and chemical potential, chiral symmetry is also dynamically broken by the self-interaction of quarks: the chiral condensate $\langle \bar q q \rangle$ is negative and large.\\
The gap equation for the quark of species $i$ relates the chiral condensates to the effective constituent quark mass:
\begin{equation}
M_i=m_i-2G\varphi_i-2K\varphi_j\varphi_k \qquad i\neq j\neq k
\label{eq:MGE} 
\end{equation}

The first term in the RHS of the equation is the bare quark mass, the second term is due to the 4-fermion interaction vertex and the third term is due to the 6-fermion interaction vertex. This vertex mixes the chiral condensates of different flavours.

\section{Fluctuations}
In many different fields the study of fluctuations can provide physical
insights into the underlying microscopic degrees of freedom. The fluctuations can
become invaluable physical observables in spite of their difficult character. Some examples of this fact are the Brownian Motion, the Cosmic microwave background and the shot noise in electric junctions.
Hence they are powerful tools to diagnose microscopic physics, to trace
back the history of the system and the nature of its elementary
degrees of freedom, as discussed in the review \cite{Asakawa} by Asakawa.\\
 The present work is focused on the fluctuations of the conserved charges in the QGP (B, Q, S) which are explored through their cumulants.
 
\section{Quark susceptibilities}
In the Grand canonical ensemble it is possible to define the fluctuation of conserved charges as:
\begin{equation}
 \chi_n\equiv \frac{1}{V}\frac{\partial^n(-\Omega/T^4)}{\partial(\mu/T)^n}=-\frac{\partial^n(\omega/T^4)}{\partial(\mu/T)^n}
\end{equation}
$\chi^n$ being the generalized susceptibility defined as the density of cumulants of order $n$. Then the kurtosis $\kappa$ and skewness $\gamma$ become:
\begin{equation}
    \kappa\sigma^2=\frac{\chi_4}{\chi_2} \qquad \frac{\gamma\sigma^3}{M}=\frac{\chi_3}{\chi_1}\qquad \sigma^2=\chi_2 \qquad M=\chi_1
\end{equation}
From a general point of view the phase-diagram of QCD is a 4-dimensional space: 3 dimensions for the quark chemical potentials and 1 for temperature.
However it is possible to choose a particular relation between the chemical potentials for reducing the phase diagram of QCD to 2 dimensions.
We perform our calculations in the following scenarios:
\begin{itemize}
\item Symmetric chemical potential: $\mu_u=\mu_d=\mu_s=\frac{1}{3}\mu_B$
\item (Quasi-)Neutral Strangeness: $\mu_u=\mu_d=\frac{1}{3}\mu_B$ , $\mu_s=0$
\item Heavy Ion Collision: $\frac{n_Q}{n_B}=0.4$, $n_s=0$
\end{itemize}
The last scenario is the one closer to the experimental conditions in HICs.
\section{Results}

\begin{figure}[h!]
	\centering
        \includegraphics[width=0.5\textwidth]{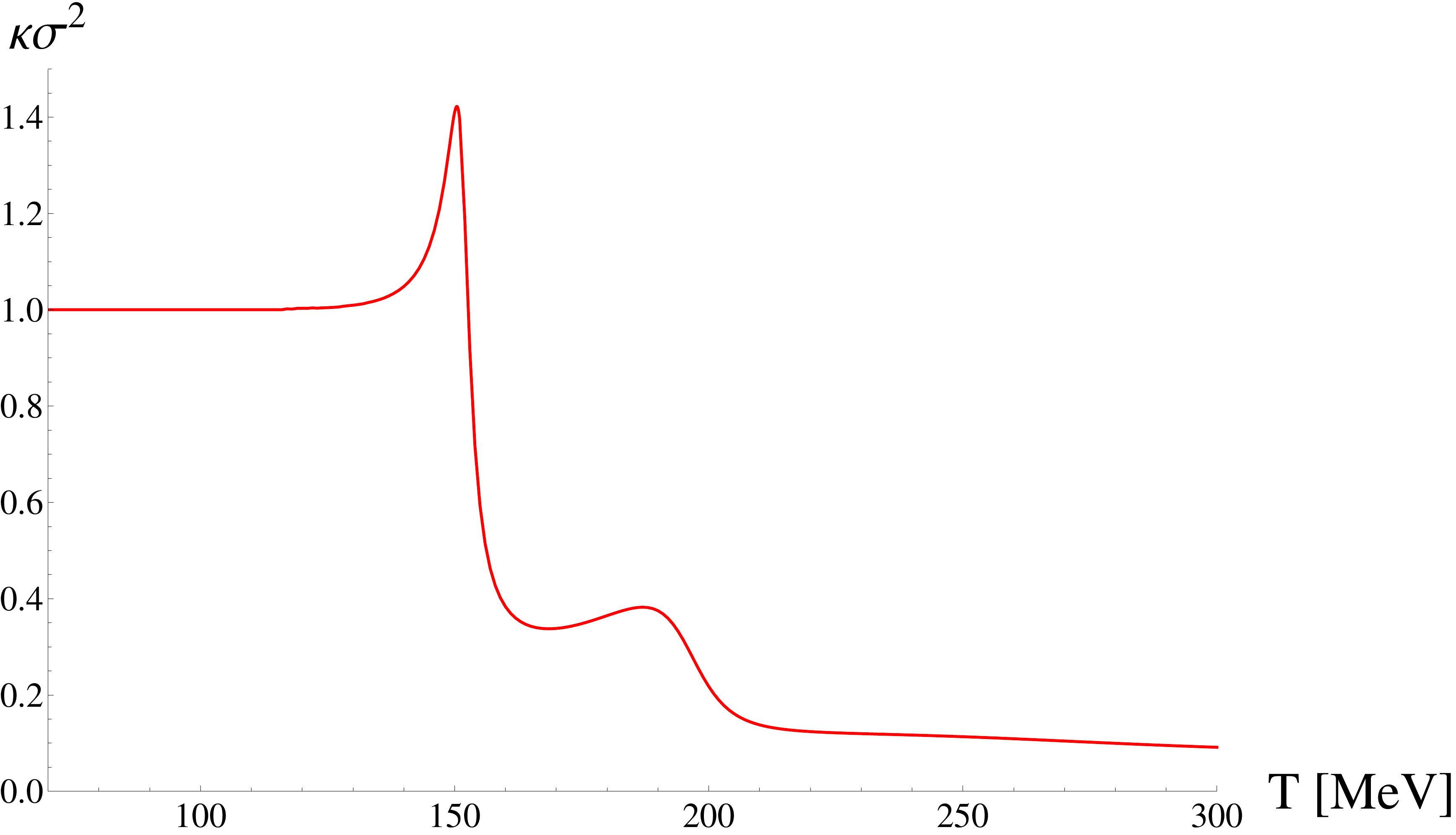}
	\caption{The ratio $\chi^4_B/\chi^2_B=\kappa\sigma^2$ at zero chemical potential.}
	\label{fig:K0}
    \end{figure}

The kurtosis at zero chemical potential (Fig.\ref{fig:K0}) as a function of temperature shows two peaks, the first one is due to the deconfinement transition and the other one is due to chiral symmetry restoration. The PNJL model, depending on the value of the parameters of the Polyakov potential $\mathcal{U}$, display two different temperatures for the chiral and deconfinement  transitions. 
The value of the kurtosis at low temperature is close to unity, above the transition it approaches $1/9$ as expected since $\kappa\sigma^2$ is proportional to the square of baryon charge carried by the elementary degrees of freedom (1 if baryons, 1/3 if quarks).
\begin{figure}[h!]
	\subfloat[][\emph{$\gamma\sigma^3/M$ as function of temperature.}]
        {\includegraphics[width=0.45\textwidth]{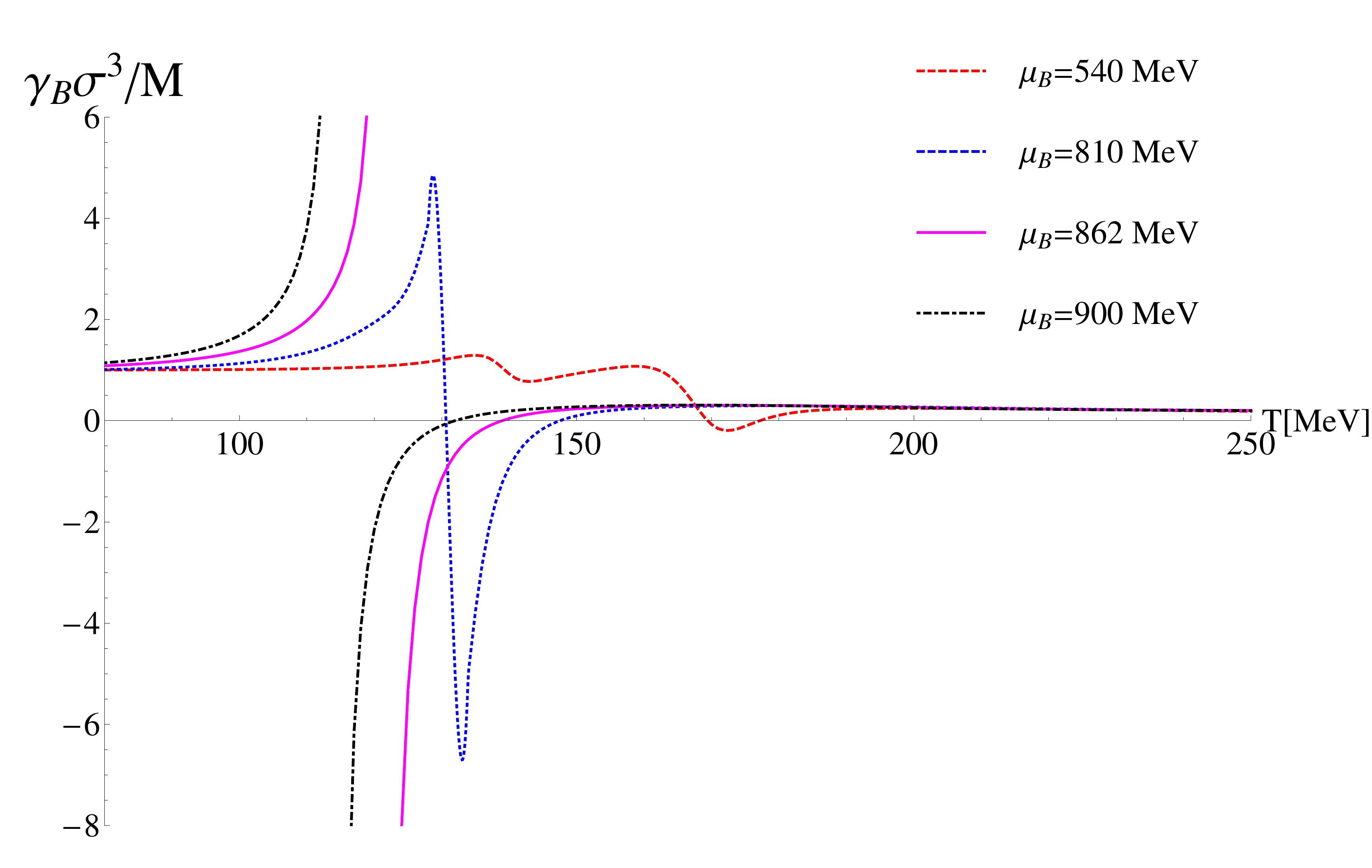}}\quad
	\subfloat[][\emph{$\kappa\sigma^2$ as a function of the temperature.}]
	{\includegraphics[width=0.45\textwidth]{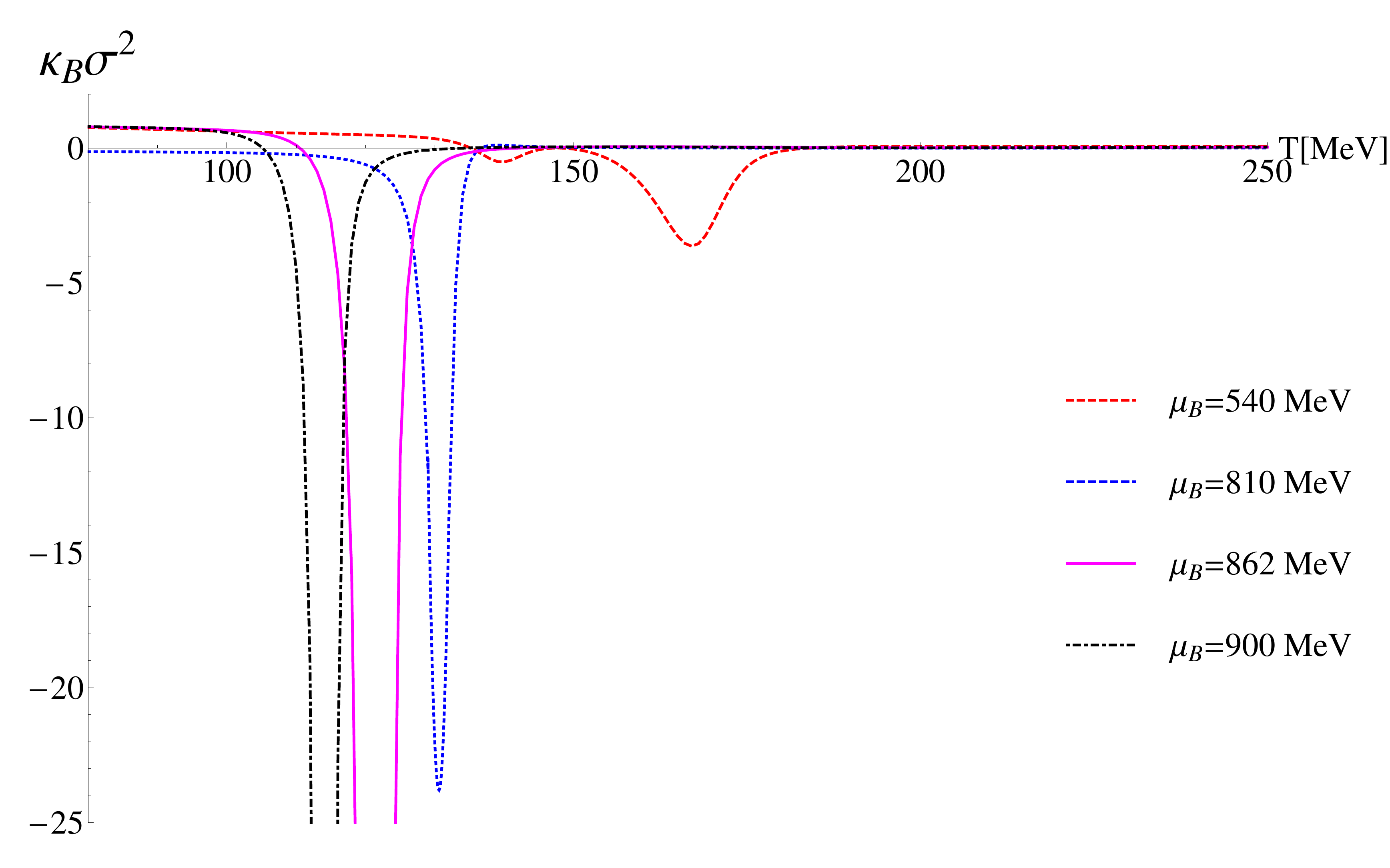}}
	\caption{Net-Baryon fluctuations as a function of the temperature for some values of chemical potential: $\mu_B=540$ MeV (dashed line),  $\mu_B=810$ MeV (dotted line), $\mu_B=862$ MeV (solid line) and  $\mu_B=900$ MeV (dash-dotted line).}  
\label{fig:KeS_mu_c}
    \end{figure}

In Fig.\ref{fig:KeS_mu_c} we display the kurtosis and the skewness at finite chemical potential. They present a decrease from low to high temperatures and this signals the deconfinement.  Fluctuations have an essential discontinuity at $\mu_B = 862$ MeV and $T=122$ MeV, in agreement with other observables.\\ 

\begin{figure}[h!]
	\subfloat[][\emph{$\gamma\sigma^3/M$ as function of baryo-chemical potential.}]
        {\includegraphics[width=0.45\textwidth]{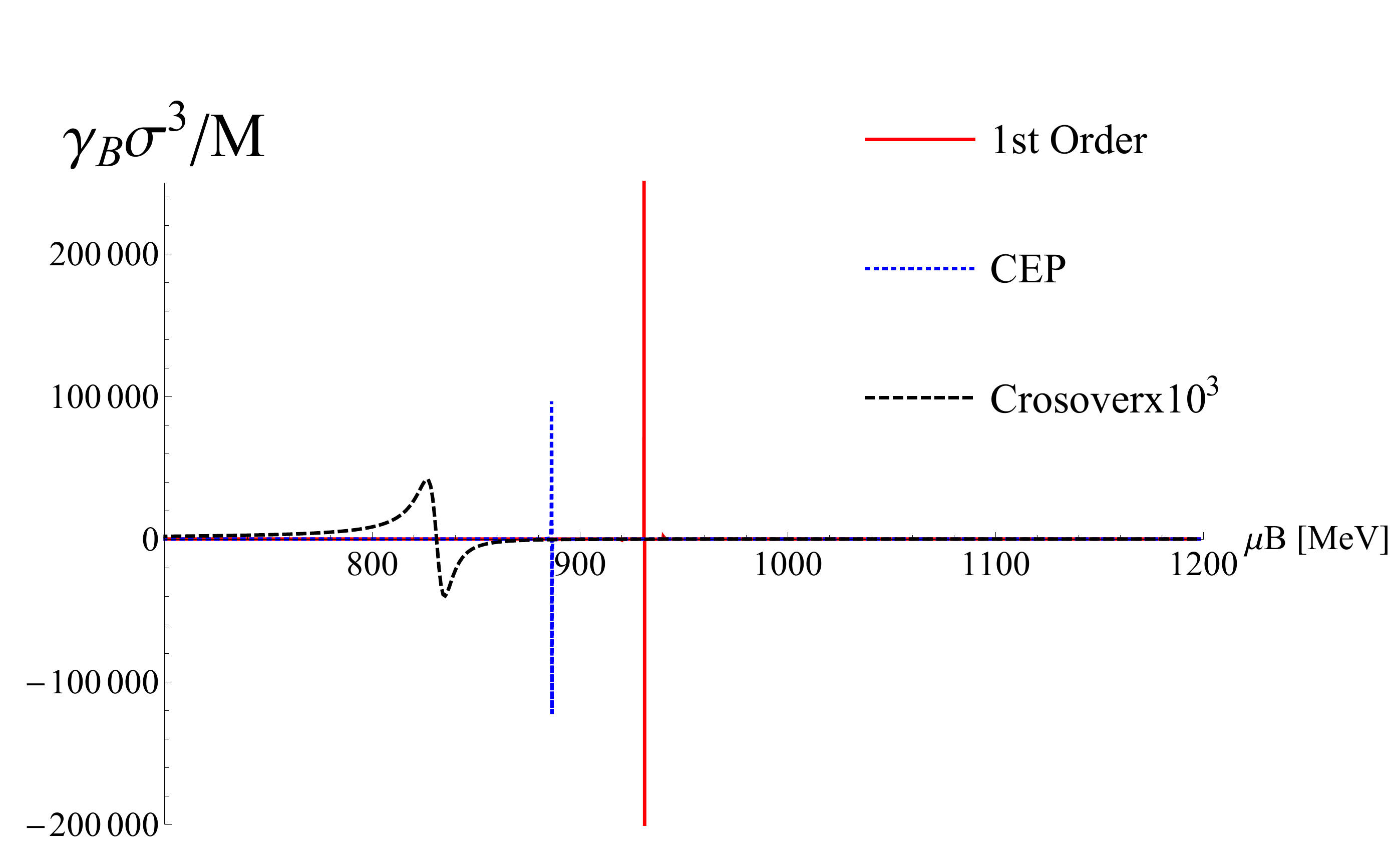}}\quad
	\subfloat[][\emph{$\kappa\sigma^2$ as function of baryo-chemical potential.}]
	{\includegraphics[width=0.45\textwidth]{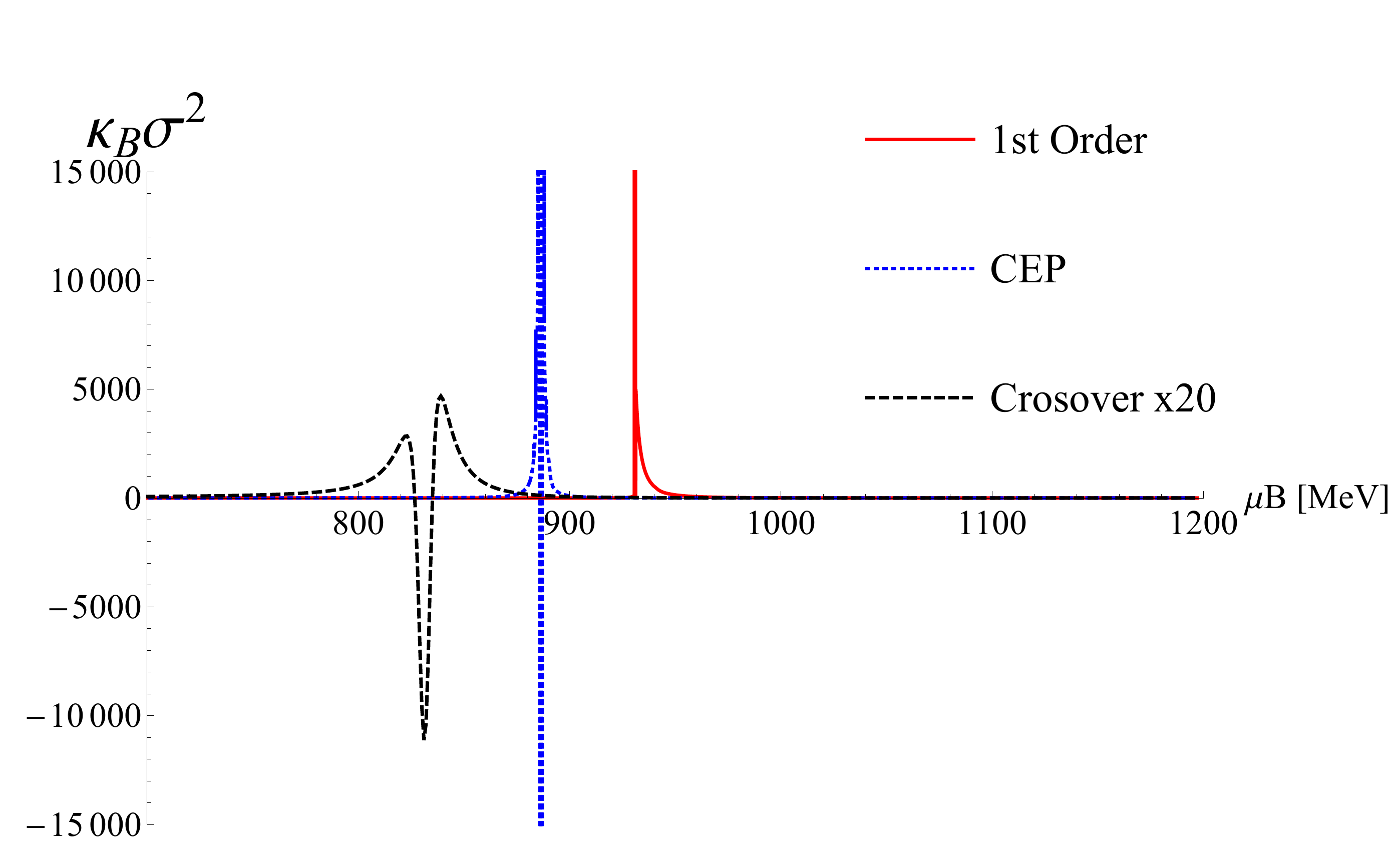}}
	\caption{T=142.9 MeV (Dashed line), T=132.9 MeV (Dotted line), T=122.9 MeV (Solid line).}
	\label{fig: KeS_T_c}
    \end{figure}
\begin{figure}[h!]
         \includegraphics[width=0.6\textwidth]{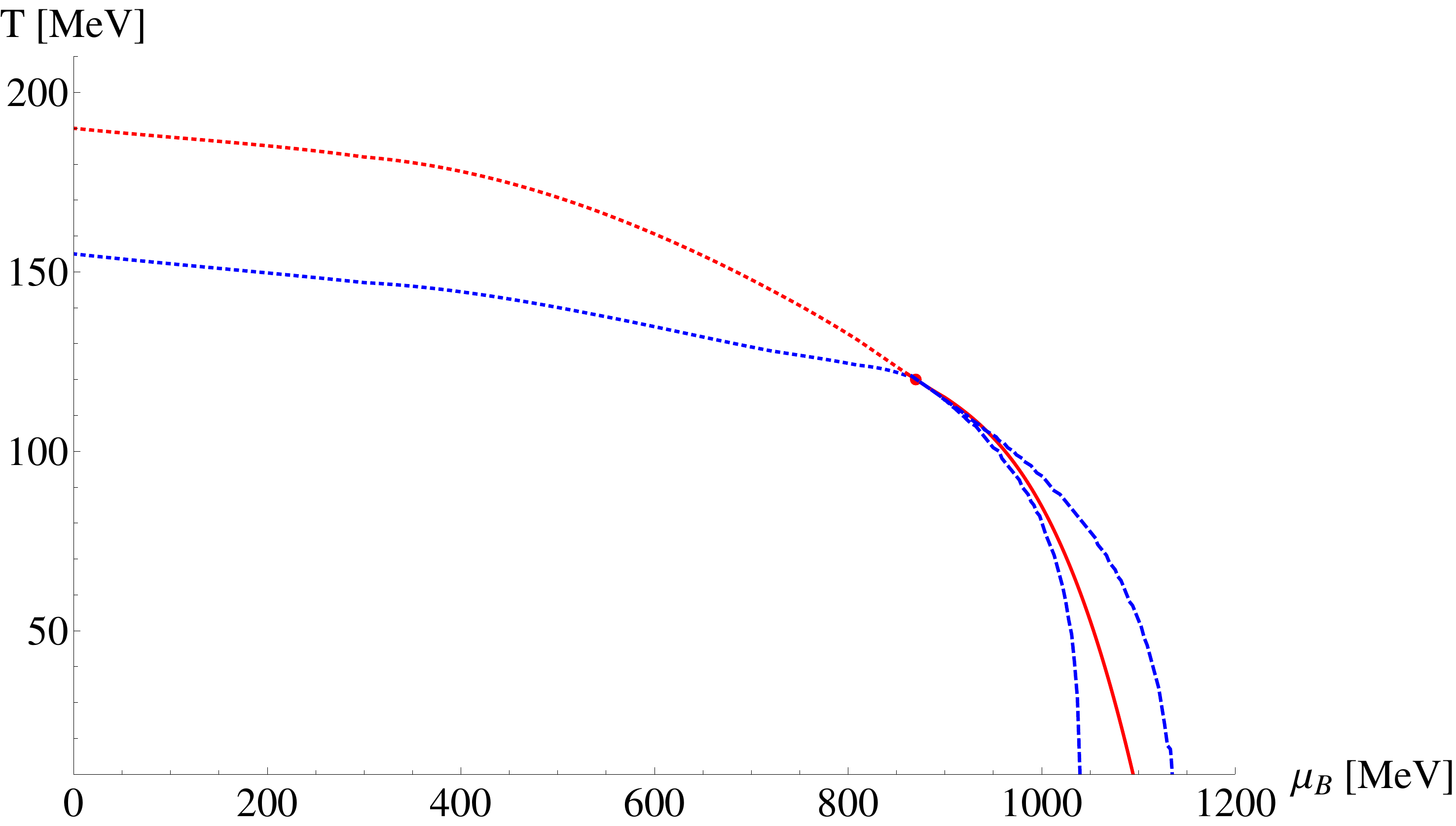}
	\caption{The phase diagram of the PNJL model. The dotted red line refers to the restoration of chiral symmetry, the dotted blue line refers to the deconfinement, the dashed blue lines represent the border of the metastable region and the solid red line corresponds to the $1^\text{st}$-order transition line.}
\label{fig:PD}
\end{figure}

In Fig. \ref{fig: KeS_T_c} we plot the kurtosis and skewness in the second scenario ($\mu_s=0$) as functions of baryo-chemical potential in three regions of the phase diagram. In the cross-over region (black line) the fluctuations are smooth and not so big. Around the CEP the fluctuations display a sizable growth. Around the $1^\text{st}$-order region the fluctuations have an essential discontinuity.\\
Near the $1^{st}$-order transition line, the Grand Canonical Potential has 2 local minima and 1 maximum. The system admits a metastable region. In Fig \ref{fig:PD} we shown the Phase Diagram of the PNJL model.

\section{Conclusions}
 In conclusion, the PNJL model seems to provide a good qualitative and semi-quantitative guidance to describe the chiral and deconfinement QCD transition. This model is not the complete theory but only an effective approach and, as such, it presents some  defects, with e.g. the deconfinement occurring at a lower temperature than the chiral transition. In spite of its limitations we plan to apply it to the study of quantities of phenomenological interest for HICs like the speed of sound, the heat capacity and fluctuations of conserved charges.\\
 
 \section*{Acknowledgments}
 This  work  was  supported  by  a University of Turin Reseach Fellowship and by STSM Grants from the COST Action CA15213 “Theory of hot matter and relativistic heavy-ion collisions” (THOR) (PC, MM, RS).

\end{document}